\documentclass[a4paper,10pt]{article}
\usepackage{cite,amsmath,amsfonts,amsthm,fullpage}
\usepackage{youngtab}
\newcommand{\Pf}{\mathop\mathrm{Pf}\nolimits}
\newcommand{\sgn}{\mathop\mathrm{sgn}\nolimits}
\newcommand{\DP}{\mathop\mathrm{DP}\nolimits}
\newcommand{\bt}{\mathbf{t}}

\theoremstyle{plain}
\newtheorem{Lemma}{Lemma}
\newtheorem{Theorem}{Theorem}

\theoremstyle{remark}
\newtheorem{Remark}{Remark}

\begin{document}
\author{J. J. C. Nimmo\thanks{Department of Mathematics, University of Glasgow,
 Glasgow G12 8QQ, UK. email: j.nimmo@maths.gla.ac.uk} \and A. Yu.
Orlov\thanks{Oceanology Institute, Nahimovskii Prospekt 36,
Moscow, Russia, email: orlovs@wave.sio.rssi.ru}}
\title{A relationship between rational and multi-soliton solutions of the
BKP hierarchy}

\maketitle

\begin{abstract}
We consider a special class of solutions of the BKP hierarchy
which we call $\tau$-functions of hypergeometric type. These are
series in Schur $Q$-functions over partitions, with coefficients
parameterised by a function of one variable $\xi$, where the
quantities $\xi(k)$, $k\in\mathbb{Z^+}$, are integrals of motion
of the BKP hierarchy. We show that this solution is, at the same
time, a infinite soliton solution of a dual BKP hierarchy, where
the variables $\xi(k)$ are now related to BKP higher times. In
particular, rational solutions of the BKP hierarchy are related to
(finite) multi-soliton solution of the dual BKP hierarchy. The
momenta of the solitons are given by the parts of partitions in
the Schur $Q$-function expansion of the $\tau$-function of
hypergeometric type. We also show that the KdV and the NLS soliton
$\tau$-functions coinside the BKP $\tau$-functions of
hypergeometric type, evaluated at special point of BKP higher
time; the variables $\xi$ (which are BKP integrals of motions)
being related to KdV and NLS higher times.
\end{abstract}

\section{Introduction}
The BKP hierarchy was introduced in \cite{DJKM,DJKM'} as a
particular reduction of the KP hierarchy of integrable equations
\cite{ZSh,DJKM}. Like the well-known KP hierarchy, the BKP
hierarchy possesses multi-soliton and rational solutions.  In
\cite{You,Nimmo}, the role of projective Schur functions
($Q$-functions) \cite{Mac} in obtaining rational solutions of the
BKP hierarchy was explained. In fact, the $Q$-functions are
polynomial $\tau$-function solutions of the BKP hierarchy Hirota
equations and these are connected to the rational solutions
through a change of dependent variables.

In \cite{q2}, certain hypergeometric series in $Q$-functions  (see
\eqref{htfBtinfty} below) were shown to be $\tau$-functions of the
BKP hierarchy. These $\tau$-functions are series of the form
\begin{equation}\label{htfq}
\tau({\bf t}_o,\xi,{\bf t}_o^*)=\sum_{\lambda \in \DP}
e^{\sum_{i=1}^\infty \xi_{\lambda_i}}
Q_\lambda(\tfrac12{\bt_o})Q_\lambda(\tfrac12{\bt_0}^*) \ ,
\end{equation}
where  $\xi=\{\xi_m: m=1, 2,\dots \}$ are arbitrary parameters,
$\xi_0=0$, $Q_\lambda$ denote projective Schur function, and the
sum is over the set $\DP$ of all partitions
$\lambda=(\lambda_1,\lambda_2,\dots,\lambda_k)$ with distinct
parts $\lambda_1>\lambda_2>\cdots>\lambda_k\ge0$. Considered as a
function of the variables $\bt_o=(t_1,t_3,\dots)$, series
\eqref{htfq} is a BKP $\tau$-function, where the set $\bt_o$ are
higher BKP times. The second set of parameters
$\bt_o^*=(t_1^*,t_3^*,\dots)$ give the evolution in a second BKP
hierarchy.

\begin{Remark}\label{1} Consider a set $S$, which
consists of distinct non-negative integers and includes zero. By
$\DP_S$ we denote the subset of all strict partitions whose parts
belong to the set $S$. By the limiting procedure: $e^{\xi_k}\to 0$
iff $k$ does not belong to the set $ S $, we can restrict the sum
in \eqref{htfq} to the set $\DP_S$. If $S$ is a finite set, then
 \eqref{htfq} is a polynomial in $\bt_o$ which describes a {\em
 rational} solution of the BKP hierarchy.
\end{Remark}

The typical choice of the BKP higher times is the following:
\begin{equation}\label{t-x}
mt_m=\sum_k^{N} \left(x_k^m-(-x_k)^m\right),\quad
mt_m^*=\sum_k^{N^*} \left(y_k^m-(-y_k)^m\right).
\end{equation}
In this case the sum in \eqref{htfq} ranges over all partitions
whose length do not exceed $k=\min(N,N^*)$.

We note that a special case of this series \eqref{htfq}, where
times were chosen as in \eqref{t-x}, and $e^{\xi_m}$ was chosen as
a step function, was considered in \cite{TW} in a combinatorial
context, not related to integrable systems.

In the present paper, we will specialize the variables $\bt^*_o$
as $\bt_\infty=(1,0,0,\dots)$, and study the $\tau$-function
\eqref{htfq} as a function of the variables $\xi_m$. We find that
series \eqref{htfq} is a multi-soliton $\tau$-function of a
different integrable hierarchy, which we call the dual BKP
hierarchy. The variables $\xi_m$ of \eqref{htfq} turn out to be
linear combinations of the time variables ${\tilde\bt}_o=({\tilde
t}_1,{\tilde t}_3,\dots )$ of the dual BKP hierarchy. We observe
that the variables $\xi_m$ (proportional to the times of the dual
BKP hierarchy) are integrals of motion of the original BKP
hierarchy and, simultaneously, the times $\bt_o$ of the original
BKP hierarchy are integrals of motion of the dual BKP hierarchy.
That is why we call these hierarchies dual to one another.

The situation we will describe is closely related to corresponding
results for the hypergeometric $\tau$-functions of the KP
hierarchy. These $\tau$-functions are described as hypergeometric
because they generalize some known hypergeometric functions of
many variables, see \cite{GR,Milne}. We note that the KP
hypergeometric $\tau$-functions yields a perturbative asymptotic
expansion for a set of known matrix integrals \cite{1,Cadiz}. They
were also used to construct new solvable matrix integrals
\cite{1',galipolli}. Other examples of hypergeometric
$\tau$-functions arise in supersymmetric gauge theories
\cite{N},\cite{LMN}, in the problem of counting of Hurwitz numbers
\cite{O},  in counting Gromov-Witten invariants of $P^1$ \cite{OP}
and in the computation of intersection numbers on Hilbert schemes
\cite{LQW}. In references \cite{NTT,T}, $\tau$-functions, which
were considered in the context of $c=1$ strings, are also of
hypergeometric type. The series for two dimensional QCD,
considered in \cite{Mig,SK}, may be related to the KP
hypergeometric $\tau$-functions also. We anticipate that
applications of similar series in $Q$-functions are found also.

The series \eqref{htfq} can be studied  in the context of random
strict partitions.  Series \eqref{htfq} generalizes the sums over
random partitions which are considered in \cite{TW}.

With regard to notation used in this paper, we will use infinite
sequences of higher times
\begin{equation}\label{tt*}
\bt =(t_1,t_2,t_3,\dots),\quad \bt^*
  =(t_1^*,t_2^*,t_3^*,\dots),
\end{equation}
\begin{equation}\label{tot*o}
\bt_o=(t_1,t_3,t_5,\dots),\quad \bt_o^*=(t_1^*,t_3^*,t_5^*\dots),
\end{equation}
and, when they appear as higher times in dual equations will be
marked with a tilde. Special cases of these, $\bt_\infty$ and
$\bt_o(q)$, will be defined in (\ref{choicetinftyB}) and
(\ref{choicetinftyqB}) below.

\section{KP and BKP $\tau$-functions}
In this section we will summarise the essential facts about
$\tau$-functions for the KP and BKP hierarchies as given in
\cite{DJKM}. The definitions of terms related to symmetric
functions may be found in \cite{Mac}.

\subsection{Schur functions as KP $\tau$-functions}
Let $A$ be the complex Clifford algebra generated by the
\emph{charged free fermions} $\psi_i$, $\psi_i^*$, where
$i\in\mathbb{Z}$ with anticommutation relations
\begin{equation}
[\psi_i,\psi_j]_+=[\psi^*_i,\psi^*_j]_+=0,\quad
[\psi_i,\psi^*_j]_+=\delta_{i,j}.
\end{equation}
Consider also the generators
\begin{equation}
    \psi(p)=\sum_{k\in\mathbb{Z}}\psi_kp^k,\quad
    \psi^*(q)=\sum_{k\in\mathbb{Z}}\psi^*_kq^{-k-1}.
\end{equation}
The vacuum expectation value is a linear functional $\langle\
\rangle\colon A\to \mathbb{C}$. For linear and quadratic elements
in $A$ it is defined by
$\langle\psi_i\rangle=\langle\psi^*_i\rangle=\langle\psi_i\psi_j\rangle
=\langle\psi^*_i\psi^*_j\rangle=0$
and
\begin{equation}
    \langle\psi_i\psi^*_j\rangle=
    \begin{cases}
      \delta_{i,j}&i<0\\
      0&i\ge0
    \end{cases},\quad
    \langle\psi_i^*\psi_j\rangle=
    \begin{cases}
      \delta_{i,j}&i\ge0\\
      0&i<0
    \end{cases}.
\end{equation}
For an arbitrary product of linear terms in $A$, Wick's Theorem
gives

\begin{equation}\label{Wick}
\langle 0|w_1 \cdots w_{2n+1}|0 \rangle =0,\quad \langle 0|w_1
\cdots w_{2n} |0\rangle =\sum_\sigma \sgn(\sigma) \langle
0|w_{\sigma(1)}w_{\sigma(2)}|0\rangle \cdots \langle 0|
w_{\sigma(2n-1)}w_{\sigma(2n)} |0\rangle ,
\end{equation}
where $w_k$ are linear terms in $A$, and $\sigma$ runs over
permutations such that $\sigma(1)<\sigma(2),\dots,
\sigma(2n-1)<\sigma(2n)$, and $\sigma(1)<\sigma(3)<\cdots
<\sigma(2n-1)$.


The connection between the anticommutation relations and the
vacuum expectation value is that $[w_1,w_2]_+=\langle
w_1w_2\rangle+\langle w_2w_1\rangle$.


For free fermion generators with $|p|\ne|q|$,
\begin{equation}
    \langle\psi(p)\psi^*(q)\rangle=\frac 1{p-q},
\end{equation}
and for higher degree products, Wick's Theorem gives
\begin{align}
    \langle\psi(p_1)\psi^*(q_1)\cdots\psi(p_n)\psi^*(q_n)\rangle
    &=\det(\langle\psi(p_i)\psi^*(q_j)\rangle)\nonumber\\
    &=\prod_i\frac{1}{p_i-q_i}
    \prod_{i<j}\frac{(p_i-p_j)(q_i-q_j)}{(p_i-q_j)(q_i-p_j)}.
\end{align}

Time evolution enters $A$ via the hamiltonian
\begin{equation}
    H(\bt)=\sum_{n=1}^\infty H_nt_n,
\end{equation}
where
\begin{equation}
    H_n=\sum_{k\in\mathbb{Z}}\psi_k\psi^*_{k+n}.
\end{equation}
For any $a\in A$, we define
\begin{equation}
  a(\bt)=e^{H(\bt)}ae^{-H(\bt)}=\exp(\mathop{\mathrm{ad}}
  H(\bt))\;a,
\end{equation}
and it may be shown that
\begin{equation}\label{eq:time evolution}
    \psi(p)(\bt)=\exp(\xi(p,\bt))\psi(p),\quad
    \psi^*(q)(\bt)=\exp(-\xi(q,\bt))\psi^*(q),
\end{equation}
where $\xi(p,\bt)=\sum_{k=1}^\infty p^kt_k$.

Consider  $g\in A$, which solves the following bilinear equation
\begin{equation}
\left[g \otimes g,\sum_{n=-\infty}^\infty \psi_n \otimes
\psi_n^*\right]=0,
\end{equation}
where the notation $[,]$ stands for the commutator, and $\otimes$
is the tensor product.  Then, one has a $\tau$-function
\begin{equation}
    \tau(\bt)=\langle g(\bt) \rangle.
\end{equation}

The simplest type of $\tau$-functions correspond to multi-soliton
solution of the KP hierarchy. Taking $g=\exp\bigl(\sum_{i=1}^n
a_i\psi(p_i)\psi^*(q_i)\bigr)$ gives the $n$-soliton
$\tau$-function
\begin{equation}\label{eq:KP n sol}
    \tau(\bt)=\det\left(\delta_{i,j}+\frac{a_i}{p_i-q_j}
    e^{\xi(p_i,\bt)-\xi(q_j,\bt)}\right) \ ,
\end{equation}
where
\begin{equation}\label{xi-KP}
\xi(p,\bt)=\sum_{m=1}^\infty p^mt_m
\end{equation}
Later, we will also need the soliton solution of the
two-dimensional Toda lattice equation (TL) \cite{UT}, which is
described by almost the same formula
\begin{equation}\label{eq:TL n sol}
    \tau(n,\bt,\bt^*)=A(\bt,\bt^*)
    \det\left(\delta_{i,j}+
    \frac{a_i}{p_i-q_j}
    e^{\xi(p_i,n,\bt,\bt^*)-\xi(q_j,n,\bt,\bt^*)}\right) \ ,
\end{equation}
where
\begin{equation}\label{xi-TL}
\xi(p,n,\bt,\bt^*)=\sum_{m=1}^\infty (p^mt_m -p^{-m}t_m^*)+n\log p
\end{equation}
and
\begin{equation}\label{TL-vacuum}
A(\bt,\bt^*)=e^{\sum_{m=1}^\infty mt_mt_m^*}
\end{equation}
which is usually omitted from the definition of the TL
$\tau$-function \cite{UT} since the transformation to nonlinear
variables removes it from the TL solution. We will also neglect
this term for the same reason. It is well-known \cite{UT}, that
any TL $\tau$-function is a $\tau$-function of the pair of KP
hierarchies with higher times respectively $\bt$ and $\bt^*$.
There exists a reduction to the one-dimensional TL, which yields
also a reduction to the nonlinear Schr\"odinger equation. This
reduction is described by the demand that the $\tau$-function of
the two-dimensional TL (up to the irrelevant factor
\eqref{TL-vacuum}) depends only on $\bt+\bt^*$. It is provided by
the condition $q_i=p_i^{-1}$ in \eqref{eq:TL n sol} and we will
use it in what follows. We shall also use the reduction to KdV,
namely the choice $q_i=-p_i$ in \eqref{eq:KP n sol}. The KdV
$\tau$-function depends only on the odd index KP higher times,
that is, on the sequence $\bt_o$.

Polynomial $\tau$-functions are obtained by considering expansions
in the parameters $p_i$ and $q_j$. First, elementary Schur
polynomials $s_i(\bt)$ are defined by
\begin{equation}
    \exp(\xi(p,\bt))=\sum_{k\ge0} s_i(\bt)p^k.
\end{equation}
Since
\begin{equation}
    1=\exp(\xi(p,\bt))\exp(-\xi(p,\bt))=
    \sum_{i\ge0}\sum_{j=0}^is_{i-j}(\bt)s_{j}(-\bt)p^i,
\end{equation}
we have the orthogonality condition
\begin{equation}
    \sum_{j=0}^i s_{i-j}(\bt)s_{j}(-\bt)=\delta_{i,0}.
\end{equation}

For all non-negative integers we can define
\begin{equation}
   s_{(a|b)}(\bt)=(-1)^b\sum_{k=0}^{b}s_{a+1+k}(\bt)s_{b-k}(-\bt)
   =(-1)^{b+1}\sum_{k=0}^{a}s_{k}(\bt)s_{a+b+1-k}(-\bt).
\end{equation}
This is the Schur function for the partition $(a+1,b^j)$, which is
written using Frobenius notation as $(a|b)$. This result is easily
proved using the Jacobi-Trudi identity.  For any partition
function written in Frobenius notation,
\begin{equation}
    s_{(a_1a_2\dots a_n|b_1b_2\dots b_n)}(\bt)=\det(s_{(a_i|b_j)}).
\end{equation}

Using this notation, \eqref{eq:time evolution} gives
\begin{equation}
    \psi_i(\bt)=\sum_{k\ge0} s_{k}(\bt)\psi_{i-k},\quad
    \psi^*_i(\bt)=\sum_{k\ge0}s_{k}(-\bt)\psi^*_{i+k}.
\end{equation}
Consequently,
\begin{equation}
    \langle\psi_i(\bt)\psi^*_j(\bt)\rangle
    =
    \sum_{k,\ell\ge0} s_k(\bt)s_\ell(-\bt)\langle\psi_{i-k}\psi^*_{j+\ell}\rangle
    =
    \sum_{k=i+1}^{i-j} s_k(\bt)s_{i-j-k}(-\bt).
\end{equation}
Hence we see that
\begin{equation}
    s_{(a|b)}(\bt)=(-1)^{b+1}\langle\psi_a(\bt)\psi^*_{-b-1}(\bt)\rangle,
\end{equation}
that is that $s_{(a|b)}$ is the KP $\tau$-function for
$g=(-1)^{b+1}\psi_a\psi^*_{-b-1}$. More generally, this shows that
an arbitrary Schur function $s_{(a_1\cdots a_n|b_1\cdots b_n)}$,
is a KP $\tau$-function for
\[
g=(-1)^{b_1+\cdots+b_n+n}\psi_{a_1}\psi^*_{-b_1-1}\cdots\psi_{a_n}\psi^*_{-b_n-1}.
\]

\subsection{$Q$-functions as BKP $\tau$-functions}
The subalgebra of $A$ invariant under the symmetry
\begin{equation}\label{eq:B symm}
  \psi_i\leftrightarrow(-1)^i\psi^*_{-i}
\end{equation}
is used in a similar way to determine BKP $\tau$-functions. There
are two bases of \emph{neutral free fermions}
\begin{equation}
    \phi_i=\frac1{\sqrt2}(\psi_i+(-1)^i\psi^*_{-i}),\quad
    \hat\phi_i=\frac i{\sqrt2}(\psi_i-(-1)^i\psi^*_{-i}),
\end{equation}
where $i\in\mathbb{Z}$, each of which generates this subalgebra.

%

Using the results for charged free fermions, the anticommutation
relations are
\begin{equation}
    [\phi_i,\phi_j]_+=[\hat\phi_i,\hat\phi_j]_+=(-1)^i\delta_{i,-j},
    \quad[\phi_i,\hat\phi_j]_+=0,
\end{equation}
and, in particular, $\phi_0^2=\hat\phi_0^2=\frac12$. Similarly,
the vacuum expectation values of quadratic elements are given by
\begin{equation}
    \langle \phi_i\phi_j \rangle=\langle \hat\phi_i\hat\phi_j \rangle=
    \begin{cases}
      (-1)^i\delta_{i,-j}&i<0\\
      \frac12\delta_{j,0}&i=0\\
      0&i>0
    \end{cases},
\end{equation}
and Wick's Theorem is used for arbitrary degree products.


The neutral free fermion generator is defined by
$\phi(p)=\sum_{n\in\mathbb{Z}}p^n\phi_n$. We have (for
$|p|\ne|p'|$)
\begin{equation}
  \langle \phi(p)\phi(p') \rangle=
  \frac12\frac{p-p'}{p+p'},
\end{equation}
and $\langle \phi(p')\phi(p) \rangle=-\langle \phi(p)\phi(p')
\rangle$. By Wick's Theorem we get
\begin{equation}
  \langle \phi(p_1)\phi(p_2)\cdots\phi(p_N) \rangle=
  \begin{cases}
    \Pf(\langle \phi(p_i)\phi(p_j) \rangle)\\
    0
  \end{cases}=\begin{cases}
    \displaystyle 2^{-N/2}\prod_{i<j}\frac{p_i-p_j}{p_i+p_j}&N\text{ even}\\
    0&\text{otherwise}
  \end{cases}.\label{eq:phi ef}
\end{equation}

The connection between the charged and neutral free fermions can
be expressed in terms of the generators as
\begin{equation}
    -q\psi(p)\psi^*(-q)+p\psi(q)\psi^*(-p)=\phi(p)\phi(q)+\hat\phi(p)\hat\phi(q).
    \label{eq:KP->BKP}
\end{equation}

In the BKP reduction, even times are set equal to zero and we
define $\bt_o =(t_1,0,t_3,0,t_5,\dots)$, and the hamiltonian
\begin{equation}
    H^B(\bt_o)=\sum_{i\ge1,\;\text{odd}}H^B_nt_n,
\end{equation}
where
\begin{equation}
    H^B_n=\frac12\sum_{i\in\mathbb{Z}}(-1)^{i+1}\phi_i\phi_{-i-n}.
\end{equation}

For the fermion generating function one has 
\begin{equation}
    \phi(p)(\bt_o)=e^{H^B(\bt_o )}\phi(p)e^{-H^B(\bt_o )}=
    e^{\hat H^B(\bt_o )}\phi(p)e^{-\hat H^B(\bt_o )}
    =e^{\xi(p,\bt_o)}\phi(p).
\end{equation}
Note also that
\begin{equation}
    H(\bt_o)=H^B(\bt_o)+\hat H^B(\bt_o),\quad [H^B(\bt_o),\hat
    H^B(\bt_o)]=0.
    \label{eq:KP->BKP H}
\end{equation}

Similar to the KP case, BKP $\tau$-functions are defined by
\begin{equation}
    \tau_B(\bt_o)=\langle h(\bt_o)\rangle,
\end{equation}
where $h$ is the Clifford algebra of the neutral free fermions
$\phi_i$. The $n$-soliton $\tau$-function is obtained by the
choice $g=\exp\bigl(\sum_{i=1}^n a_i\phi(p_{i})\phi(q_{i})\bigr)$.

The Schur $q$ polynomials are defined by
\begin{equation}
    \exp(2\xi(p,\bt_o))=\sum_{k\ge0}q_k(\bt_o)p^k.
\end{equation}
Thus
\begin{equation}
    \phi_i(\bt_o)=\sum_{k\ge0}q_k(\tfrac12\bt_o)\phi_{i-k}.
\end{equation}
We have
\begin{equation}\label{eq: phi q_i,j}
    \langle\phi_i(\bt_o)\phi_j(\bt_o)\rangle=
    \frac12q_i(\tfrac12\bt_o)q_j(\tfrac12\bt_o)+
    \sum_{k=1}^j(-1)^{k}q_{k+i}(\tfrac12\bt_o)q_{j-k}(\tfrac12\bt_o).
\end{equation}

Since
\begin{equation}
    1=\exp(2\xi(p,\bt_o))\exp(-2\xi(p,\bt_o))=\sum_{i,j}q_i(\bt_o)q_{j-i}(-\bt_o)=
    \sum_{i,j}(-1)^{i-j}q_i(\bt_o)q_{j-i}(\bt_o)p^j,
\end{equation}
for all $n>0$ we have
\begin{equation}\label{eq:B othog}
    \sum_{i=0}^n(-1)^{i}q_i(\bt_o)q_{n-i}(\bt_o)=0.
\end{equation}
This is trivial if $n$ is odd and if $n=2m$ is even then it gives
\begin{equation}
    q_m(\bt_0)^2+2\sum_{k=1}^m(-1)^{k}q_{m+k}(\bt_o)q_{m-k}(\bt_o)=0.
\end{equation}

We can also define
\begin{equation}\label{eq: q_a,b}
    q_{a,b}(\bt_o)=q_a(\bt_o)q_b(\bt_o)+2\sum_{k=1}^b(-1)^k
    q_{a+k}(\bt_o)q_{b-k}(\bt_o).
\end{equation}
If follows from the orthogonality condition \eqref{eq:B othog}
that
\begin{equation}
    q_{a,b}(\bt_o)=-q_{b,a}(\bt_o),
\end{equation}
and in particular, $q_{a,a}(\bt_o)=0$. Comparing \eqref{eq: phi
q_i,j} and \eqref{eq: q_a,b}, it is clear that
\begin{equation}
    q_{a,b}(\tfrac12\bt_o)=2\langle\phi_a(\bt_o)\phi_b(\bt_o)\rangle.
\end{equation}

Now consider $\lambda=(\lambda_1,\lambda_2,\dots,\lambda_{2n})$
where $\lambda_1>\lambda_2>\cdots\lambda_{2n-1}>\lambda_{2n}\ge0$.
Note that this is a partition with an extra trivial part 0
included if necessary to ensure that the number of parts is even.
The set of such strict, or distinct part, partitions is denoted
$\DP$. For $\lambda\in\DP$ we define
\begin{equation}
    Q_\lambda(\tfrac12\bt_o)
    =\Pf(q_{\lambda_i,\lambda_j}(\tfrac12\bt_o)).
\end{equation}
This is the Schur $Q$-function. By Wick's theorem,
\[
    Q_\lambda(\tfrac12\bt_o)
    =\Pf(2\langle\phi_{\lambda_i}(\bt_o)\phi_{\lambda_j}(\bt_o)\rangle)
    =2^n\langle\phi_{\lambda_1}(\bt_o)
    \phi_{\lambda_2}(\bt_o)\cdots\phi_{\lambda_{2n}}(\bt_o)\rangle.
\]

\subsection{Hypergeometric $\tau$-functions}
These $\tau$-functions were introduced by one of the authors in
the KP case \cite{hypsol} and the BKP case \cite{q2}.

In the KP case, let $r$ be a function of one variable and for any
partition $\lambda$, let $r_\lambda(x)=\prod_{(i,j)\in\lambda}
r(x-i+j)$, the product being over all vertices in the Young
diagram.
Then
\begin{equation}
    \tau(n,\bt,\bt^*)=\sum_\lambda
    r_\lambda(n)s_\lambda(\bt)s_\lambda(\bt^*)
\end{equation}
where  $\bt =(t_1,t_2,t_3,\dots),\ \bt^*
  =(t_1^*,t_2^*,t_3^*,\dots)$, is the KP hypergeometric $\tau$-function.

In the BKP case, $r_\lambda$ has a different definition: if
$\lambda=(\lambda_1,\dots,\lambda_k)$ then
$r_\lambda=\prod_{i=1}^k r(1)r(2)\cdots r(\lambda_i)$. If we
introduce new variables $ r(k)=e^{\xi_k-\xi_{k-1}},\quad
\xi_{-1}=0$ then
$r_{\lambda}=\prod_{i=1}^{\ell(\lambda)}e^{\xi_{\lambda_i}} $

With these definitions,
\begin{equation}\label{eq:BKP hyp}
   \tau(\bt_o,\xi,\bt^*_o)=\sum_{\lambda\in\DP}2^{-\ell(\lambda)}r_\lambda
   Q_\lambda(\tfrac12\bt_o)Q_\lambda(\tfrac12\bt^*_o)=
   \sum_{\lambda\in\DP}2^{-\ell(\lambda)}
   \prod_{i=1}^{\ell(\lambda)}e^{\xi_{\lambda_i}}
   Q_\lambda(\tfrac12\bt_o)Q_\lambda(\tfrac12\bt^*_o)
\end{equation}
is the BKP hypergeometric $\tau$-function. By Remark~\ref{1} one
can restrict sum (\ref{eq:BKP hyp}) to the sum which ranges over
$\DP_S$.

It can be show that this \emph{is} a $\tau$-function since
\begin{equation}\label{fermtauB}
\tau (\bt_o ,\xi,\bt^*_o)= \langle
0|e^{\sum_{i\ge1,\;\text{odd}}H^B_nt_n}
e^{\sum_{n=-\infty}^{\infty}(-)^{n}\xi_n:\phi_n \phi_{-n}:}
e^{\sum_{i\ge1,\;\text{odd}}H^B_{-n}t_n^*}|0\rangle
\end{equation}

\section{Main results}

\subsection{KP infinite soliton solution}
Let
\begin{equation}\label{g-KP}
g^{\rm sol}=\exp\left(\sum_{i,j\ge0}a_{i,j}\psi(p_i)\psi^*(q_j)\right)
\end{equation}
Then
\begin{align}
    \tau^{\rm sol}:&=\langle
    g^{\rm sol}({\bt})\rangle \label{solitons-KP}\\
    &=1+\sum_{i,j}a_{i,j}\langle \psi(p_i)\psi^*(q_j)({\bt})\rangle+
    \sum_{i,j,k,l}(a_{i,j}a_{k,l}-a_{i,l}a_{j,k})
    \langle \psi(p_i)\psi^*(q_j)\psi(p_k)\psi^*(q_l)({\bt}) \rangle
    +\cdots\nonumber\\
    &=1+\sum_{i,j}{i \choose j}\frac{1}{p_i-q_j}e^{\xi(p_i,\bt)-
    \xi(q_j,\bt)}\nonumber\\
    &\qquad+
    \sum_{i,j,k,l}{i,j\choose k,l}\frac{1}{(p_i-q_j)(p_k-q_l)}
    \frac{(p_i-p_k)(q_j-q_l)}{(p_i-q_l)(q_j-p_k)}
    e^{\xi(p_i,\bt)-\xi(q_j,\bt)+\xi(p_k,\bt)-\xi(q_l,\bt)}+\cdots, \nonumber
\end{align}
where
\begin{equation}
    {i,j\choose k,l},
\end{equation}
denotes the $2\times2$ minor of the infinite matrix $(a_{ij})$
containing the $i$th and $j$th rows and the $k$th and $l$th
columns.

If $a_{i,j}=s_{(i|j)}(\bt^*)$ then the coefficients can be written
as
\begin{equation}
    s_\lambda(\bt^*)
\end{equation}
for all partitions $\lambda$.

\subsection{KdV soliton solution}

 The KdV reduction of the KP soliton solution \eqref{eq:KP n sol},
\begin{equation}\label{g-KdV}
g^{\rm sol}_{\rm
KdV}=\exp\left(\sum_{i\ge0}a_{i}p_i\psi(p_i)\psi^*(-p_i)\right)\ ,
\end{equation}
gives rise to the following soliton $\tau$-function, which depends
only on the higher KP times with odd numbers
$\bt_o=(t_1,t_3,t_5,\dots)$,
\begin{align}\label{KdV-sol}
    &\tau^{\rm sol}_{\rm KdV}:=\langle
    g^{\rm sol}_{\rm KdV}(\bt_o)\rangle=
    \det\left(\delta_{i,j}+\frac{a_ip_i}{p_i+p_j}
    e^{\xi(p_i,\bt_o)-\xi(-p_i,\bt_o)}\right)
    \\
    &=1+\sum_{i}\frac{1}{2}e^{\eta_i}+
    \sum_{i>j}\frac{1}{2^2}
\frac{(p_i-p_j)^2}{(p_i+p_j)^2}e^{\eta_i+\eta_j}\nonumber\\
    &\qquad+
    \sum_{i>j>k}\frac{1}{2^3}
    \frac{(p_i-p_j)^2(p_i-p_k)^2(p_j-p_k)^2}{(p_i+p_j)^2(p_i+p_k)^2(p_j+p_k)^2}
    e^{\eta_i+\eta_j+\eta_k}+\cdots,
    \nonumber
  \end{align}
where
\begin{equation}
\eta_i=2\sum_{m=1}^\infty\ { t}_{2m-1} p_i^{2m-1}+\log a_i \ ,
\quad i=0,1,2,\dots
\end{equation}

\begin{Remark} We note that the fractional linear
transformation of the (complex) plane of spectral parameters
$p_i\in {\mathbb C}$
\begin{equation}\label{linearfrac}
p_i \to \frac{ap_i+b}{cp_i+d} \ ,\quad \frac{ad-bc}{ad+bc}=\pm 1\
,\quad i=1,2,\dots
\end{equation}
leaves invariant the factors
\begin{equation}
\frac{(p_i-p_j)^2}{(p_i+p_j)^2}\ ,\quad i,j=1,2,\dots.
\end{equation}
It is easy to see that there are two distinct types of invariance;
$a=d=0$, $b,c\ne0$ and $a,d\ne0$, $b=c=0$.
\end{Remark}

\subsection{NLS and one-dimensional Toda lattice soliton $\tau$-function}
Let us consider the reduction of the TL soliton $\tau$-function
\eqref{eq:TL n sol} to the one-dimensional Toda lattice (1DTL)
reduction, which is $q_i=p_i^{-1}$, see \cite{UT}. If, in
addition, $|p_i|=1$, then, it is also a reduction to the nonlinear
Schr\"odinger equation (NLS). For the multi-soliton tau function
we have
\begin{equation}
g^{\rm sol}_{\rm 1DTL}=\exp\left(\sum_{i\ge 0} \frac
{a_i}{2}(p_i-p_i^{-1})\psi(p_i)\psi^*(p_i^{-1})\right)
\end{equation}
\begin{align}
    &\tau^{\rm sol}_{\rm 1DTL}(n,{\bt},{\bt}^*)=
    \tau^{\rm sol}_{\rm 1DTL}(n,{\bt}+{\bt}^*):=\nonumber\\
    &=\langle
    g^{\rm sol}_{\rm 1DTL}(n,{\bt},{\bt}^*)\rangle=
    \det\left(\delta_{i,j}+\frac{a_i(p_i-p_i^{-1})}{2(p_i-p_j^{-1})}
    e^{\xi(p_i,n,{\bt},{\bt}^*)-
    \xi(p_j^{-1},n,{\bt},{\bt}^*)}\right)\nonumber\\
    &=1+\sum_{i}\frac{1}{2}e^{\eta_i}+
    \sum_{i>j}\frac{1}{2^2}
\frac{(p_i-p_j)^2}{(p_ip_j-1)^2}e^{\eta_i
    +\eta_j}\nonumber\\
    &\qquad+
    \sum_{i>j>k}\frac{1}{2^3}
    \frac{(p_i-p_j)^2(p_i-p_k)^2(p_j-p_k)^2}
    {(p_ip_j-1)^2(p_ip_k-1)^2(p_jp_k-1)^2}    e^{\eta_i+\eta_j+\eta_k}+
    \cdots,\label{sol-1DTL}
  \end{align}
  where $\bt =(t_1,t_2,t_3,\dots),\ \bt^*
  =(t_1^*,t_2^*,t_3^*,\dots)$, and
\begin{equation}\label{xi-1DTL}
\eta_i=\sum_{m=1}^\infty
(p_i^m-p_i^{-m})({ t}_m +{ t}_m^*)+2n\log p_i+\log a_i \ ,\quad
i=0,1,2,\dots
\end{equation}
For the nonlinear Schr\"odinger equation the $n$-dependence of the
$\tau$-function is irrelevant.
\begin{Remark} Here, the fractional linear
transformations
\begin{equation}
p_i \to \pm \frac{ap_i+b}{bp_i+a} \ ,\quad i=0,1,2,\dots
\end{equation}
where $a$ and $b$ are not both zero, leave invariant the factors
\begin{equation}
\frac{(p_i-p_j)^2}{(p_ip_j-1)^2} \ , \quad i,j=0,1,2,\dots.
\end{equation}
\end{Remark}

\subsection{BKP infinite soliton solution}
Now writing $q_i=-p_i$ and choosing skew-symmetric matrix entries
$a_{ji}=-a_{ij}$ in \eqref{g-KP} gives
\begin{equation}
g^{\rm sol}=\exp\left(\sum_{0\le
i<j}a_{i,j}\bigl(-p_j\psi(p_i)\psi^*(-p_j)+p_i\psi(p_j)\psi^*(-p_i)\bigr)\right).
\end{equation}
By \eqref{eq:KP->BKP} this may be rewritten as
\begin{equation}
g^{\rm sol}=\exp\left(\sum_{0\le
i<j}a_{i,j}\bigl(\phi(p_i)\phi(p_j)+\hat\phi(p_i)\hat\phi(p_j)\bigr)\right).
\end{equation}
Since $[\phi_i,\hat\phi_j]_+=0$,
$[\phi_i\phi_k,\hat\phi_j\hat\phi_l]=0$ and so we can factorize as
$g=h\hat h$ where
\begin{equation}
 h=\exp\left(\sum_{i<j}a_{i,j}\phi(p_i)\phi(p_j)\right),\quad
 \hat
 h=\exp\left(\sum_{i<j}a_{i,j}\hat\phi(p_i)\hat\phi(p_j)\right).
\end{equation}

Then, we have
\begin{align}
    \tau_{\rm B}^{\rm sol}(\bt_o):&=\langle
    h(\bt_o)\rangle=\langle
    \hat h(\bt_o)\rangle\nonumber\\
    &=1+\sum_{0\le i<j}a_{i,j}\langle \phi(p_i)\phi(p_j)(\bt_o)\rangle\nonumber\\
    &\qquad+\sum_{0\le i<j<k<l}(a_{i,j}a_{k,l}-a_{i,k}a_{j,l}+a_{i,l}a_{j,k})
    \langle \phi(p_i)\phi(p_j)\phi(p_k)\phi(p_l)(\bt_o) \rangle+\cdots\nonumber\\
    &=1+\sum_{0\le i<j}(i,j)\frac12\frac{p_i-p_j}{p_i+p_j}
    e^{\xi(p_i,\bt_o)+\xi(p_j,\bt_o)}
    \nonumber\\
    &\qquad+\sum_{0\le i<j<k<l}(i,j,k,l)\frac1{2^2}
    \frac{(p_i-p_j)(p_k-p_l)}{(p_i+p_j)(p_k+p_l)}e^{\xi(p_i,\bt_o)
    +\xi(p_j,\bt_o)+\xi(p_k,\bt_o)+\xi(p_l,\bt_o)}+\cdots,\label{eq:tau_B}
\end{align}
where
\begin{equation}
    (i,j,k,l),
\end{equation}
denotes the pfaffian minor of the infinite skew-symmetric matrix
$(a_{ij})$ containing the $i$th, $j$th, $k$th and $l$th lines.
Using \eqref{eq:KP->BKP H} gives
\begin{equation}
    \tau^{\rm sol}(\bt_o)=\tau_B^{\rm sol}(\bt_o)^2.
\end{equation}

If $a_{i,j}=q_{i,j}(\tfrac12\bt_o^*)$ then the coefficients can be
written as
\begin{equation}
    Q_\lambda(\tfrac12\bt_o^*)
\end{equation}
for all partitions $\lambda$ into distinct parts.

\begin{Remark}\label{rem:frac-lin}
The factors
$$
(i,j)\frac{p_i-p_j}{p_i+p_j}, \quad (i,j,k,l)
\frac{(p_i-p_j)(p_k-p_l)}{(p_i+p_j)(p_k+p_l)}, \ \dots
$$
 in (\ref{eq:tau_B}), are invariant under the transformation
\begin{equation}
 p_i \to \frac{ap_i+b}{cp_i+d} \ ,\quad i=0,1,2,\dots ,
\end{equation}
\begin{equation}\label{BKPa=gammaq}
 a_{i,j}\to \gamma a_{i,j} \ ,\quad i,j=0,1,2,\dots ,
\end{equation}
where $a,b,c,d$ are any complex numbers such that
\begin{equation}\label{BKPgamma}
\frac{ad+bc}{ad-bc}=\gamma \ ,
\end{equation}
is not zero or infinity.
\end{Remark}

\begin{Remark}\label{rem:p=0}
Although the free fermion generator is not defined if its
parameter is 0, i.e. $\phi(0)$ does not make sense, the limit
\begin{equation}
    \lim_{p'\to0}\langle \phi(p)\phi(p')\rangle=\frac12
\end{equation}
as given by \eqref{eq:phi ef}, is well defined.
\end{Remark}

\subsection{Useful Lemma}
Let us introduce the following notation:

\begin{equation}\label{choicetinftyB}
{\bf t}_\infty=(1,0,0,0,\dots) \ ,
\end{equation}
and
\begin{equation}\label{choicetinftyqB}
{\bf t}_o(q)=(t_1(q),t_3(q),t_5(q),\dots ) \ ,\quad
t_{2m-1}(q)=\frac{2}{(2m-1)(1-q^{2m-1})}\ ,
 \quad t_{2m}=0, \quad m=1,2,\dots
\end{equation}
\begin{Remark}\label{R}
Let us notice that $\bt_\infty$ can be viewed as given by
(\ref{t-x}), where we take $x_1=x_2=\cdots=x_N=N^{-1}$ and
$N\to\infty$. Similarly, ${\bf t}_o(q)$ is given by (\ref{t-x}),
where $x_k=q^{k-1},k=1,2,\dots$.
 As for ${\bf
t}_\infty$, if $f$ satisfies
$f(ct_1,c^3t_3,c^5t_5,\dots)=c^df(t_1,t_3,t_5,\dots)$ for some
$d\in {\mathbb Z}$, we have $\hbar^df({\bf t}_o(q))\to f({\bf
t}_\infty) $ as $\hbar:=\log q \to0$. In that sense
(\ref{choicetinftyB}) may be considered as a limit of
(\ref{choicetinftyqB}) as $q\to 1$.

\end{Remark}
We have

\begin{Lemma}Let $\lambda=(\lambda_1,\dots,\lambda_k)$ be a
strict partition. Then
\begin{equation}\label{stinftyB}
Q_{\lambda}(\tfrac12\bt_\infty)=
\prod_{i=1}^k \frac {1}{\lambda_i!}\prod^k_{i<j}\frac{\lambda_i-\lambda_j}
{\lambda_i+\lambda_j} \ ,
\end{equation}
 and
\begin{equation}\label{Qinftyq'}
Q_{\lambda}\left(\tfrac12\bt_o(q)\right)= \prod_{i<
j}^k\frac{q^{\lambda_i}-q^{\lambda_j}} {q^{\lambda_j+\lambda_i}-1}
\prod_{i=1}^k\frac {(-q;q)_{\lambda_i}}{(q;q)_{\lambda_i}}\ ,
\end{equation}
where
\begin{equation}\label{q-pochhammer}
(p;q)_m :=(1-p)(1-pq)\cdots (1-pq^{m-1})
\end{equation}
\end{Lemma}

\subsection{Hypergeometric functions related to the projective
Schur functions}

 We consider hypergeometric $\tau$-functions
(\ref{htfq}), where we specialize the variable ${\bf t}_o^*$
respectively by (\ref{choicetinftyB}) and (\ref{choicetinftyqB}):
\begin{equation}\label{htfBtinfty}
\tau ({\bf t}_o ,\xi,{\bf t}_\infty  )= \sum_{{\lambda}\in
\DP_S}\frac{Q_{\lambda
}\left(\tfrac12\bt_o\right)}{2^{\ell(\lambda)}{H^*_\lambda}}
\prod_{i=1}^{\ell(\lambda)}e^{\xi_{n_i}} ,
\end{equation}
\begin{equation}\label{htft(q)}
\tau ({\bf t}_o ,\xi,{\bf t}_o(q)  ) = \sum_{{\lambda}\in
\DP_S}\frac{Q_{\lambda}\left(\tfrac12\bt_o\right)}
{2^{\ell(\lambda)}{H^*_\lambda(q)} }
\prod_{i=1}^{\ell(\lambda)}e^{\xi_{n_i}}
 ,
\end{equation}
 where
\begin{equation}\label{hooks-prod}
H^*_\lambda=Q_{\lambda }\left(\tfrac12\bt_\infty\right)^{-1},\quad
H^*_\lambda(q)= Q_{\lambda }\left(\tfrac12\bt_o(q)\right)^{-1}
\end{equation}
are respectively so-called product-of-shifted-hook-length
\cite{Mac}, which generalize the notion of the factorial for
strict partitions and its $q$-analog (shifted hook polynomial). We
took into account Remark~\ref{1}, to restrict sums over all strict
partitions to the subset $\DP_S$.

In the case that $\DP_S$ is the set of all strict partitions,
namely, $\DP$, the series \eqref{htfBtinfty} and \eqref{htft(q)}
may be considered as multi-variable generalization of
hypergeometric function (respectively, basic hypergeometric
function), which we obtain when $\ell(\lambda)=1$ and $\bt_o$ is
of form \eqref{t-x} where $N_1=1$.

The notation $Q(x^{(N)})$ below will be used for $Q(\frac{\bf t_o}{2})$, where
$\bt_o$ is defined by (\ref{t-x}).
Let all parameters $b_k$ be not equal to negative integers.
 Let in (\ref{htfBtinfty}) we choose
\begin{equation}\label{E3'}
e^{\xi_n}=  \frac{\prod_{i=1}^p \Gamma (a_i+n)\Gamma
(a_i)^{-1}}{\prod_{i=1}^s\Gamma(b_i+n)\Gamma(b_i)^{-1} }=
\frac{\prod_{i=1}^p (a_i)_n}{\prod_{i=1}^s(b_i)_n}
\end{equation}
Then tau function (\ref{htfBtinfty}) defines the following
hypergeometric function
\begin{equation}\label{pFs}
{}_{p}F_s(a_1,\dots,a_p;b_1,\dots,b_s; x^{(N)}) :
=\sum_{\lambda\in \DP \atop \ell(\lambda\le N}^\infty
2^{-\ell(\lambda)} \frac{\prod_{k=1}^{p}(a_k)_{\lambda} }
{\prod_{k=1}^{s}(b_k)_{\lambda} }\frac
{Q_\lambda(x^{(N)})}{H^*_\lambda} \ ,
\end{equation}
which generalizes the hypergeometric function of one variable
\begin{equation}\label{onevarE3'}
{}_{p}F_s(a_1,\dots,a_p;b_1,\dots,b_s; x^{(1)}) =\sum_{n=0}^\infty
\frac{\prod_{k=1}^{p}(a_k)_{n} } {\prod_{k=1}^{s}(b_k)_{n} }\frac
{x^n}{n!}
\end{equation}

The function (\ref{pFs}) was introduced in \cite{q2}. Here we
introduce the $q$-deformed version of (\ref{pFs}).
If in (\ref{htft(q)}) we choose
\begin{equation}\label{E3'q}
e^{\xi_n}=  \frac{\prod_{i=1}^p
(q^{a_i};q)_n}{\prod_{i=1}^s(q^{b_i};q)_n} \ ,
\end{equation}
tau function (\ref{htft(q)}) defines the hypergeometric function
\begin{equation}\label{qpFs}
{}_{p}\Phi_s(a_1,\dots,a_p;b_1,\dots,b_s; x^{(N)}):
=\sum_{\lambda\in \DP \atop \ell(\lambda\le N}^\infty
2^{-\ell(\lambda)} \frac{\prod_{k=1}^{p}(q^{a_k};q)_{\lambda} }
{\prod_{k=1}^{s}(q^{b_k};q)_{\lambda} }\frac
{Q_\lambda(x^{(N)})}{H^*_\lambda(q)} \ ,
\end{equation}
which generalizes the basic hypergeometric function of one variable
\begin{equation}
{}_{p}F_s(a_1,\dots,a_p;b_1,\dots,b_s; x^{(1)}) =\sum_{n=0}^\infty
\frac{\prod_{k=1}^{p}(q^{a_k};q)_{n} }
{\prod_{k=1}^{s}(q^{b_k};q)_{n} }\frac {x^n}{(q;q)_n}
\end{equation}
Hypergeometric functions (\ref{pFs}) and (\ref{qpFs}) may be also
considered as multisoliton tau functions, see next subsection.

\subsection{Soliton solutions and rational solutions}

Let us recall, that in Remark~\ref{1} the set of the partitions
$\DP_S$ was defined via a set of distinct non-negative integers
$S$ which includes zero.

\begin{Theorem} Let $ \tau({\bf t}_o,\xi,{\bf t}_\infty)$ be defined by
\eqref{htfBtinfty}, and $\tau^{\rm sol}(\tilde{{\bf
t}},\tilde{{\bf t}}^*)$ be defined by \eqref{eq:tau_B}, where
\begin{equation}
 p_m=\frac{am+b}{cm+d} \ ,\quad m \in S
\end{equation}
(in particular, one can take integer momentum $p_m=m$), with any
$a,b,c,d$ such that
\begin{equation}\label{BKPabcd}
\frac{ad+bc}{ad-bc}=\gamma \ ,
\end{equation}
is not zero or infinity, and
\begin{equation}\label{BKPa=q}
 a_{i,j}=\gamma q_{i,j}(\tfrac12\bt_o^*)\
\end{equation}
 Let for a given set of the numbers $p_m$, $m \in S $, the
variables $\ \tilde{{\bf t}},\tilde{{\bf t}}^*$ are related to the
variables $\xi$ as
\begin{equation}\label{ThTm}
\xi_{m}=\sum_{k=1}^\infty (p_m^k \tilde{t}_k
-p_m^{-k}\tilde{t}^*_k)+\log  m!.
\end{equation}
Then we have
\begin{equation}\label{ThA}
\tau^{\rm sol}(\tilde{{\bf t}})=\tau({\bf t}_o,\xi,{\bf
t}_\infty).
\end{equation}
\end{Theorem}
\begin{proof}
Let us compare \eqref{htfBtinfty} and \eqref{eq:tau_B}. First
replace $\bt_o$ with $\tilde\bt_o$ in \eqref{eq:tau_B}. Then a
typical term on the right hand side is
\begin{equation}
    (n_1,n_2,\dots,n_k)\;2^{-k}\prod_{i<j}
    \frac{p_{n_i}-p_{n_j}}{p_{n_i}+p_{n_j}}
    \prod_{i=1}^{k}e^{\xi(p_{n_i},\tilde\bt_o)},
\end{equation}
where $0\le n_1<n_2<\cdots<n_k$ and $k$ is even. We can set
$\xi(p_k,\tilde\bt_o)=\xi_k$, choose the parameters $p_k=k$ and
the pfaffian elements to be (for $i<j$)
\begin{equation}\label{eq:pf el}
(i,j)=\begin{cases}
 \dfrac{2q_{j}(\tfrac12\bt_o)}{j!}&i=0\\[6pt]
 \dfrac{q_{j,i}(\tfrac12\bt_o)}{i!\;j!}&i>0
\end{cases},
\end{equation}
so that
\begin{equation}\label{eq:pf}
    (n_1,n_2,\dots,n_k)=\begin{cases}
    2\prod_{i=1}^k\frac1{n_k!}\;Q_\lambda(\tfrac12\bt_o)&n_1=0\\
    \prod_{i=1}^k\frac1{n_k!}\;Q_\lambda(\tfrac12\bt_o)&n_1>0
    \end{cases}
\end{equation}
where $\lambda$ is the partition into distinct parts
$(n_k,n_{k-1},\dots,n_1)$.

Thus the typical term may be written as
\begin{equation}
2^{-\ell(\lambda)}\prod_{i=1}^{\ell(\lambda)}e^{\xi_k}
Q_\lambda(\tfrac12\bt_o) Q_\lambda(\tfrac12\bt_\infty),
\end{equation}
for any partition into distinct parts. The partitions into an odd
number of distinct parts come from those terms for which $n_1=0$.
The extra factors of 2 in the pfaffian element in \eqref{eq:pf el}
and \eqref{eq:pf} are needed because in the case $n_1=0$, the
length of the partition $\ell(\lambda)$ is $k-1$ not $k$.

This establishes the connection between \eqref{htfBtinfty} and
\eqref{eq:tau_B}.
\end{proof}

The hypergeometric function (\ref{pFs}) is an example of
multi-soliton tau function of the dual BKP hierarchy, evaluated at
 special values of times ${\tilde \bt}_o$, see (\ref{E3'}) and (\ref{ThTm}).

\begin{Theorem}
Tau functions  $\tau ({\bf t}_o ,\xi,{\bf t}_\infty  )$ and
$\tau^{\rm sol}_{\rm KdV}(\tilde{{\bf t}}_o)$ are defined
respectively
 by \eqref{htfBtinfty} and by \eqref{KdV-sol}. Let us choose
$p_m$ in \eqref{KdV-sol} by
\begin{equation}
 p_m=\frac{am+b}{cm+d} \ ,\quad m \in S \ ,
\end{equation}
where $a,b,c,d$ are any parameters with the property
\begin{equation}
\frac{ad-bc}{ad+bc}=\pm 1
\end{equation}
 (in particular, one can choose integer
momentum $p_m=m, \ m\in S$). Let the numbers $\xi_m$ in
(\ref{htfBtinfty}) be related to $\eta({\bt}_o,p_m)$ in
\eqref{KdV-sol} by
\begin{equation}\label{kdvxit}
\xi_{m}-\log m!=\eta_m  :=2\sum_{k=1}^\infty
p_m^{2k-1} \tilde{t}_{2k-1}
\end{equation}
Then
\begin{equation}\label{Th-B}
\tau^{\rm sol}_{\rm KdV}(\tilde{{\bf t}}_o)=\tau({\bf
t}_\infty,\xi,{\bf t}_\infty).
\end{equation}
\end{Theorem}

The hypergeometric function (\ref{pFs}), where
$x_1=x_2=\cdots=N^{-1}$ and  $N\to \infty$, is an example of
multi-soliton KdV tau function, evaluated at special values of
times ${\tilde \bt}_o$, , see (\ref{E3'}), (\ref{kdvxit}) and
Remark~\ref{R}.

\begin{Theorem}
Tau functions $\tau ({\bf t}_o ,\xi,{\bf t}_o(q) )$ and $\tau^{\rm
sol}_{\rm 1DTL}(\tilde{n},\tilde{{\bf t}},\tilde{{\bf t}}^*)$ are
defined respectively by \eqref{htft(q)} and  by \eqref{sol-1DTL}.
Let $p_m$ in \eqref{sol-1DTL} be chosen by
\begin{equation}
p_m=\pm \frac{aq^m+b}{bq^m+a} \ ,\quad m \in S
\end{equation}
(in particular, $p_m=q^{m},\ m\in S $). Let the numbers $\xi_m$ in
(\ref{htft(q)}) be related to $\eta({\bt}_o,p_m)$ in
\eqref{sol-1DTL} by
\begin{equation}\label{1DTLxit}
\xi_{m}-\log \frac{(q;q)_{m}}{(-q;q)_{m}}=\eta_m
 :=2\sum_{k=1}^\infty (p_m^k-p_m^{-k})(
\tilde{t}_k+\tilde{t}^*_k) +2{\tilde n}\log p_m
\end{equation}
Then
\begin{equation} \label{Th-C}
\tau^{\rm sol}_{\rm 1DTL}(\tilde{n}, \tilde{{\bf t}}+\tilde{{\bf
t}}^*) =\tau({\bf t}_o(q),\xi,{\bf t}_o(q)).
\end{equation}
For the particular choice $p_m=m,\ m\in S$, this is also NLS
multi-soliton tau function.
\end{Theorem}

The hypergeometric function (\ref{qpFs}), where
$x_k=q^{k-1},k=1,2,\dots$ and $N\to\infty$, is an example of
multi-soliton 1DTL tau function evaluated at special values of
times ${\tilde \bt}+{\tilde \bt}^*$, see (\ref{E3'q}),
(\ref{1DTLxit}) and Remark~\ref{R}.

\begin{Remark}
In case $S$ is a finite set, the polynomial $\tau$-function of
type \eqref{htfBtinfty} (and \eqref{htft(q)}) is related to the
soliton $\tau$-function with a finite number of solitons.
\end{Remark}
\begin{Remark} We note that the higher times ${\bf t}_o$ of the BKP
hierarchy we started with are integrals of motion for (solitonic)
$\tau$-function \eqref{eq:tau_B} of the second BTL hierarchy.
Simultaneously, the higher times $ {\tilde n},\tilde{{\bf
t}},\tilde{{\bf t}}^*$ play the role of integrals of motion for
the original BKP hierarchy. We therefore call these hierarchies
dual to each other.
\end{Remark}
\begin{Remark}
An $\infty$-soliton solution with spectral parameters lying on a
lattice appeared in  \cite{BL,LS1,LS2} in a different way and in a
different context. Other links between soliton and rational
solutions of the KP hierarchy were found in \cite{Mir}.
\end{Remark}

\section{Discussion}
An interesting problem is to study the asymptotic behaviour of
hypergeometric $\tau$-functions. We hope to apply methods of
soliton theory to conduct this study. We note that the asymptotic
behaviour of infinite soliton $\tau$-functions, similar to those
considered in the present paper, was studied in \cite{LS1}.

We hope to apply the series \eqref{htfq} to certain problems.

(1) Let us consider an integral
\begin{equation}\label{intSchurB''}
I(N,{\bf t}_o ,{\bf t}^*_o )=\frac {1}{N!}\int\int_\Gamma \cdots
\int\int_\Gamma \prod_{i<j}^N\frac{(z_i-z_j)}{(z_i+z_j)}
\frac{(z_i^*-z_j^*)}{(z_i^*+z_j^*)} \prod_{k=1}^N
e^{\sum_{n=1,3,\dots}^\infty
\left(z^n_kt_n+{z^*_k}^nt^*_n\right)}\mu(z_k{z^*}_k)dz_kdz^*_k,
\end{equation}
where $\Gamma$ is a integration domain in each $(z_k,z^*_k)$ plane
($k=1,\dots,N$), and $\mu$ is a function such that
\begin{equation}\label{domainGammaB'}
\int \int_\Gamma \mu (zz^*)z^n dzdz^*=\int \int_\Gamma \mu(zz^*)
{z^*}^n dzdz^*=\delta_{n,0},
\end{equation}
and
\begin{equation}\label{intzz^*B'}
\int \int_\Gamma \mu (zz^*)z^n{z^*}^m
dzdz^*=2\delta_{n,m}e^{\xi_n}.
\end{equation}

The series \eqref{htfq} (in the case that the sum ranges over
partitions of length $\ell(\lambda)\le N$) yields the asymptotic
expansion for the integral \eqref{intSchurB''} \cite{q2}
\begin{equation}\label{inttauB'}
I(N,{\bf t}_o ,{\bf t}^*_o )=\sum_{\lambda \in \DP ,l(\lambda)\le
N}2^{-l({\lambda})}
e^{\sum_{i=1}^{\ell(\lambda)}\xi_{\lambda_i}}Q_{\lambda
}(\tfrac12{\bf t}_o) Q_{\lambda }(\tfrac12{\bf t}_o^*).
\end{equation}
The restriction $\ell(\lambda)\le N$ makes the difference between
the r.h.s. of \eqref{inttauB'} and $\tau(\bt_o,\xi,\bt_o^*)$.

In the limit $N\to \infty$ (which is typical for applications),
the restriction $\ell(\lambda)\le N$ is irrelevant for the
perturbation series related to \eqref{intSchurB''}, and therefore,
this series coincides with \eqref{htfq}. Also, in the case that at
least one of the sets ${\bf t}_o,{\bf t}_o^*$ has the form of
\begin{equation}\label{x/-x'}
mt_m=\sum_k^{N} \left(x_k^m-(-x_k)^m\right),\quad
mt_m^*=\sum_k^{N} \left(y_k^m-(-y_k)^m\right),
\end{equation}
 then by the bosonization
formulae and Wick's Theorem we get that
\begin{equation}\label{ell>M'}
 Q_\lambda(\tfrac12
{\bf t}_o)=0 \ , \quad \ell(\lambda)>N
\end{equation}
Therefore, in this case, the integral $I (N,{\bf t}_o ,{\bf t}^*_o
)$ is the BKP $\tau$-function $\tau(\bt_o,\xi,\bt_o^*)$.

It may be interesting to apply \eqref{htfq} to matrix models and
to statistical models where partition functions reduce to
integrals \eqref{intSchurB''}, see, for instance, \cite{Kostov}
for examples of integrals similar to \eqref{intSchurB''}.

It is interesting to compare integrals \eqref{intSchurB''} with
supersymmetric matrix models \cite{Guhr}.

(2) The series \eqref{htfq} can be studied also in the context of
random (strict) partitions.

Random  strict partitions were considered in \cite{TW} and, in
particular, the ``shifted'' measure $Q_\lambda(x)Q_\lambda(y)$  on
(strict) partitions, were considered in \cite{TW}. In this paper,
the series \eqref{htfq}, where all $\xi_n=0$, and where
$\lambda_1$ does not exceed a certain given number was studied.

Let us remark that the expression
\begin{equation}\label{plancherel}
e^{\sum_{i=1}^\infty \xi_{\lambda_i}} Q_\lambda({\bf
t}_\infty)Q_\lambda({\bf t}_\infty)=\prod_{i=1}^{\ell(\lambda)}
e^{\xi_i} \left(\prod_{i=1}^k \frac
{1}{\lambda_i!}\prod^k_{i<j}\frac{\lambda_i-\lambda_j}
{\lambda_i+\lambda_j}\right)^2=\left(\frac{1}{H^*_\lambda}\right)^2
e^{\sum_{i=1}^\infty \xi_{\lambda_i}},
\end{equation}
(where $H^*_\lambda$ is known to be related to the number of
shifted tableau of the shape $\lambda$, see \cite{Str},\cite{Mac})
in the case
\begin{equation}
\xi_n =0,\ n=0,1,2,\dots
\end{equation}
may be considered as the analogue of the Plancherel measure
\cite{Olshanski}, while in the case
\begin{equation}
e^{\xi_n} =(z)_n(1-z)_n \ , \quad
(z)_n=\frac{\Gamma(z+n)}{\Gamma(z)} \ ,\quad \ n=0,1,2,\dots
\end{equation}
as an analog of the so-called $(z)$-measure on random partitions
(see \cite{Olshanski}).

(3) Finally, we note that KdV soliton solutions with integer
momenta was first considered in \cite{BL}. Each KdV solution of
this type yields a wave operator $\partial_t^2-\Delta_{2n-1} -u$,
where the potential $u=2\partial^2_{t_1}\tau^{\rm sol}_{\rm
KdV}(t_1)$, which satisfies the so-called generalized Huygens
principle.

\section*{Acknowledgements}
The authors are grateful to T. Shiota for useful discussion. The
work was supported in part by the Russian Foundation for
Fundamental Researches (Grant No 02-02-17382), and the Program of
Russian Academy of Science ``Mathematical Methods in Nonlinear
Dynamics". We thank Claire Gilson and Chris Athorne for the
organization of the visit of one of the authors (A.O.) to the
University of Glasgow, during which work on this paper was
initiated.

\end{document}